# Why echo chambers form and network interventions fail: Selection outpaces influence in dynamic networks[1].

*Christian Steglich*

*Institute for Analytical Sociology (IAS)*
*Department of Management and Engineering*
*Linköping University*
*Norra Grytsgatan 10*
*SE-601 74 Norrköping, Sweden*

*AND*

*Interuniversity Centre for Social Science Theory and Methodology (ICS)*
*Department of Sociology*
*University of Groningen*
*Grote Rozenstraat 31*
*NL-9712TG Groningen, The Netherlands*

*christian.steglich@liu.se*

*Tel. +46 700 896 671*

[1] The research reported in this paper benefitted from discussions with Lorien Jasny, Fabian Held, the colleagues in the 'ecco' group at the IAS Norrköping, and in the research clusters 'statistics and social network analysis' and 'norms & networks' at the ICS Groningen. The author's work at the ICS was partly funded by NWO (grants 411-12-027 and 406-13-017). The work at IAS has received funding from the ERC (grant 324233), Riksbankens Jubileumsfond (grant DNR M12-0301:1), and the Swedish Research Council (grants DNR 445-2013-7681 and DNR 340-2013-5460).

## Abstract

Are online networking services complicit in facilitating social change for the worse? In two empirically informed simulation studies, we give a proof-of-concept that the speed of networking and the amplification of network actors' relational preferences can have a profound impact on diffusion dynamics on social networks, essentially counteracting the benefits that should accrue from networking according to the *strength of weak ties* argument. Our findings can help understand variations in homogeneity of network neighbourhoods, i.e., in the degree to which these neighbourhoods act as "echo chambers", as well as the high context-dependency of success rates for a certain type of network intervention studies. They suggest that the general facilitation of connectivity like it today happens on the internet, combined with the use of personalisation algorithms, has strong and insufficiently understood effects on dynamic processes unfolding on the affected social networks.

## Introduction

The rise of computer-facilitated connectedness has had two immediate and undisputed effects on our social lives. First, social networking sites, instant messaging platforms and message board communities today make it possible to get into contact with a much larger set of people than we ever could have contacted offline. Second, the time and effort needed for the initiation (or termination) of such online connections nowadays is minuscule compared to what is needed for managing offline contacts. If we define *network speed* as the intensity at which opportunities for changing one's personal network arise (i.e., the number of opportunities for adding or dropping contact persons, per time unit), we can say that online-facilitated social networks show an *increased network speed* compared to traditional offline social networks. This speed increase is so large that resource-constrained humans quickly reach the limit of what they can meaningfully process[1], even after factoring in the reduced costs of online relationship management. We will show that this first major difference between online and offline networking, innocuous as it may seem, has unanticipated social consequences that go beyond how people's networks look like. But there is another major difference between online and offline networking that we want to study.

For resource constrained humans facing this increased speed of networking opportunities in their online activities, the question arises how to optimally allocate their limited cognitive resources (Davenport & Beck, 2001; Tufekci, 2013). One way of coping with this challenge is the reliance on technical support offered by the providers of online networking infrastructure. Based on a user's past online behaviour, these providers' learning algorithms make an educated guess of what the user's preferences may have been. Subsequently, the user's online experience is *personalised*[2] by filtering new content based on rankings according to these inferred preferences. The net effect of these "filter bubbles" (Pariser, 2011) is one of increased probabilities of encountering (and choosing) preferred opportunities, i.e., one of *amplified preferences*[3] in users' choices. This amplification of preferences is the other major difference between (current[4]) online networking and the offline case that we diagnose.

In this paper, we investigate what the two major departures of online from offline social networks identified above generally imply for inter-individual processes that unfold on the affected network. Concretely, we show that identical processes of information diffusion, contagion and peer influence will produce vastly different outcomes when the network on which they occur changes at *different speeds*, and with *differentially amplified preferences*. On a conceptual level and by way of social simulation, we investigate the combined effects of these dimensions in two studies that mimic prototypical topics of social network research. The first topic is the formation of so-called *echo chambers*, i.e., the segregation of a diverse, heterogeneous population into local network neighbourhoods that consist of homogeneous individuals. The second topic is the utilisation of social networks as diffusion channels in so-called *peer-led interventions*. These target a few individuals at central positions in a network to promote a change of behaviour or attitude. Because of the targets' centrality, the promoted change subsequently diffuses efficiently through the whole network by way of inter-personal influence.

At first sight, the two topics may appear as having little in common. While the concept of echo chambers is clearly relevant for understanding online communities, the closest analogue to peer-led interventions in online networks may be found in literature on *viral marketing*. However, in terms of the conceptual backbone needed for understanding the two topics, in terms of their underlying *network mechanisms*, they are very close: Both address situations in which an actor variable and a relationship variable co-evolve, and where accordingly an approach that accounts for this co-evolution is called for. As we will show, in both applications, the speed of rewiring the network as well as actors' preference strength when doing the rewiring, are key to understanding the emergent outcome of interest. It is therefore expedient to treat both topics in the unifying, domain-independent framework provided by *network science* (Brandes et al., 2013). Doing so, we hope to contribute to a solution of domain-specific puzzles by showcasing them as instances of more general principles of how networks behave as interactive systems. And we hope to contribute to a deeper understanding of the effects of online connectedness.

The specific outcomes we will investigate are, on the one hand, the *homogeneity* of local network neighbourhoods (i.e., its *echo-chamberness*) and, on the other hand, the *success* of a network intervention (i.e., the degree to which the intervention produces the desired outcome). In the following, we will briefly introduce the specific research topics and sketch the puzzles applied researchers are facing. We then describe an empirical network study that will be used as reference scenario to inform our two simulation studies, which follow afterwards. The investigated outcomes turn out to be highly sensitive to the two dimensions of networking identified above. We finish with a discussion of what these findings mean in their specific research context, and a brief reflection on the social consequences of providing networking infrastructure and algorithms.

**The case of echo chambers**

As understood here, an *echo chamber* stands shorthand for a (potentially online) community of like-minded individuals who almost exclusively interact with one another and are otherwise sheltered from information and from others that could challenge their beliefs. The notion was coined by Cass Sunstein (2001) in an essay based on the experience of increasing polarisation of the US public sphere during the Clinton impeachment procedure and the election of George W. Bush as president, and has recently been invoked as explanatory framework for understanding the British vote to leave the European Union, and the election of Donald Trump as US president. It remains a scientifically unresolved question whether such phenomena nowadays are decisively fuelled by online infrastructure provided in the shape of social networking platforms and search engines, as some have forcefully argued (DiMaggio et al., 2004; Farrell, 2012). Recently, commissioned (Bakshy et al., 2015) as well as independent researchers (Benkler et al., 2017; Brühl et al., 2017) have argued that effects of online algorithms and infrastructure on opinion homogeneity in individuals' network neighbourhoods (i.e., the degree to which individuals find themselves in echo-chambers) exist, but are dwarfed by the effect of these individuals' unamplified preferences. While this has failed to convince the critics (Pariser, 2015; Sandvig, 2015; Tufekci, 2015), it is true and has been known for long that also in the absence of the internet, humans do sort themselves into homogeneous groups (Homans, 1950; Lazarsfeld & Merton, 1954; McPherson et al., 2001).

> *Insert Textbox 1 about here* <

What we want to add to this discussion is that, few exceptions notwithstanding (Boxell et al., 2017), the studies giving the all-clear to the internet companies may be trapped in their own filter bubble because they tend to exclusively rely on the empirical analysis of online-collected data (Bakshy et al., 2015; Benkler et al., 2017; Brühl et al., 2017). As such, they are not in a position to make any statements about how their online scenarios compare to what would be the case in a world unaffected by online infrastructure. It is a *counterfactual* study of this type that we aim to contribute with Study 1 in this paper. Working with a reference scenario that can be interpreted as "uncontaminated by the internet", we show that the above-diagnosed internet-typical departures from the offline situation actually do generate more homogeneous network neighbourhoods, i.e., echo chambers, than we would otherwise observe. The reason for this is that the network dynamics strongly determine the input of peer influence: by selectively picking a certain type of actors as social contacts, these actors will wield higher influence over the individual than other types of actors would if they were selected as social contacts instead[5]. We think that in natural offline social interaction, humans have developed a balance between how they react to the input they receive from peers, and how they adjust the composition of their peer group. With online infrastructure having a strong impact on the latter, this balance is jeopardised. The increased network speed online makes it more likely that attention given to any particular social contact will be redirected to somebody else, and hence has the general effect of levelling out differences between others' influences on any particular actor. This dwindling interpersonal influence in the traditional sense is replaced by influence of the larger pool of contact candidates that personalisation algorithms selectively help us pick from. In other words, online rewiring can either steer us into an echo chamber or pull us out of one, depending on what the providers of online infrastructure have in mind for us (see box).

**The case of network interventions**

The parallel occurrence of (and balance between) the network change process (selection) and the process of individual adjustment to network contacts (influence) is a key aspect in the above reasoning. When contact rewiring algorithms increasingly determine which information sources we are exposed to, by suggesting which news to read and which people to get into contact with, these algorithms are increasingly picking for us the sources that will influence us. In consequence, classical forms of peer influence get undermined. In particular, those peers that offline would have been influential over a considerable time period can online quickly be replaced by whatever the rewiring algorithms deem to be fitting *influencers* (a term now used for, e.g., the stars of the popular YouTube product promotion channels). It is this undermining of offline peer influence that will be addressed in detail in Study 2 of this paper, again taking a counterfactual approach in which the effect of offline-influentials is assessed under various scenarios, departing from an empirically observed offline situation.

Information about who is influential in offline networks can be found in the literature on (offline[6]) peer-led network interventions (Valente, 2005; 2012; Watts & Dodds, 2007). In such an intervention, researchers aim to exploit the occurrence of peer influence processes for propagating behaviour or attitude change, often a public health message and corresponding behaviour (e.g., *safe sex practices*, Latkin, 1998; *not to take up smoking*, Campbell et al., 2008; *drinking more water*; Smit et al., 2016). The procedure is to first identify influential opinion leaders in the target group (Starkey et al., 2009; Valente & Pumpuang, 2007) and specifically administer the intervention to this group of *influentials* only, thus reducing intervention costs compared to a treatment of the whole target group. Typically, the socially best-connected top 15% according to sociometric popularity indicators constitute this group of influentials (Kelly & Stevenson, 1995, cf. Campbell et al., 2008). If these influentials (a) accept the training, if they (b) indeed are, and stay, influential also after the training, and if (c) the message and behaviour change they propagate indeed diffuses through the network, then the intervention can be successful.

> *Insert Textbox 2 about here* <

Studies reviewing such interventions reveal that empirical success rates fluctuate a lot (see box). We suspect that one reason for this varying success is insufficient consideration of our two key dimensions, network speed and preference amplification, when designing the interventions. Our reasoning is the flipside of the echo chamber reasoning above: we think that network speed and the expression of preferences in decisions about network change are actually stronger in empirical settings than what diffusion researchers typically assume, namely, a static network[7]. When a network is not changing at all, the well-connected actors who were diagnosed as influential in the beginning will stay well-connected and hence influential over time. The more network turnover there is, the more likely they will be replaced by other, initially not influential actors. The stronger the tendency to pick similar others in this network turnover process (*homophily*; Lazarsfeld & Merton, 1954), the more the initially influential actors will be sorted into echo chambers of similar actors, whom they cannot influence further because they already are similar. The actors that still could be influenced by them get, in turn, locked into their own echo chambers of similar others. This way, network speed and the strength of homophily (i.e., similarity-seeking) in the network change process together produce an effect of insulating actors from dissimilar peers, who are the only ones who could still exert a peer effect leading to a change in opinion or behaviour.

So again we have an effect of network speed (static vs. changing at different rates) as well as an effect of homophily-based selection and de-selection (network responsiveness to the process unfolding on the network). The dimensions identified above as crucial departures of online from offline networking thus acquire a slightly different meaning in this second setting. As Study 2 will show, the systemic responsiveness of networked systems to act as efficient diffusion channels is strongly reduced if the system is allowed to respond to the diffusion by faster and more selective rewiring of the networks. Like travellers avoiding a location where a disease outbreak has been signalled, actors who disconnect themselves from behaviourally dissimilar others will reduce the likelihood of new behaviour spreading through the whole network. In the case of peer-led interventions, this counteracts the purpose of these interventions. Failure to consider such rewiring might help explain the differences in success rates of such interventions that have been reported in the literature.

After having introduced both cases, it is worth pointing out that the same two dimensions help us gain a deeper understanding of two topics and issues normally seen as quite separate: the formation of echo chambers, and the success rates of peer-led interventions. Our approach of treating both in the same formal framework showcases the benefit of generalisation that is characteristic for network science.

**Empirical anchoring**

The simulation studies will build on *stochastic actor-based network modelling* (Snijders, 1996; Snijders et al., 2010), with more than 300 published research articles probably the most widely applied family of network models for statistical inference based on empirically observed dynamic network data. More precisely, we will make use of the network-behaviour co-evolution variant of this model family (Steglich et al., 2006; 2010). These models express the joint dynamics of a network and individual behaviour[8] of the network actors as a Markov process in continuous time. In a nutshell, the model consists of four interrelated components: (1) an exponential model of actor-specific waiting times for opportunities to change their network, interpretable as the speed of network change; (2) a conditional logit discrete choice model for actors to change their network contacts, reflecting the combined effects of preferences and constraints for connecting to specific others; (3) an exponential model of actor-specific waiting times for opportunities to change their behaviour, interpretable the speed of behaviour change; and (4) a conditional logit discrete choice model for actors to change their own behaviour, reflecting their behavioural preferences. For more details about the model family, we refer the interested reader to the cited reference papers. Because of the complexity of network data and the resulting intractability of the likelihood function, these models require simulation-based parameter estimation (Snijders, 2001; Snijders et al., 2007). We will exploit these simulation facilities by using the models as a data-generating tool under various parameter sets (see also Steglich, 2007; Schaefer et al., 2013; Snijders & Steglich, 2015; Lakon et al., 2015).

**Obtaining realistic parameters**

In order to get an empirically meaningful initial set of parameters for simulating the co-evolution of a network with a behaviour dimension, we estimated a stochastic actor-based model for one of the schools in the ASSIST study (Campbell et al., 2008). The actors in this empirical data set are 245 adolescents of one school cohort aged 12-13 years at the baseline assessment. We obtained our estimates for the time period between the one- and two-year follow-up assessments. At these assessments, students were asked to nominate up to six friends (below simply called *the network*) and report their current smoking behaviour, which we recoded to an ordinal four-point scale (below simply called *the behaviour*). Descriptives of the larger study are given elsewhere (Campbell et al., 2008; Mercken et al., 2012). The estimates of the actor-based model are shown in Table 1.

> *Insert Table 1 about here* <

What the estimates mean is, first, that the empirical rate of change is more than twice as high for the network as for the behaviour, which stresses the relevance of our point that network rewiring is an important intervening mechanism with potentially far-reaching consequences if one wants to study behaviour change unfolding on a network. The other parameters indicate revealed network-constrained preferences when taking decisions about whom to select and de-select (upper part of the table) and how strongly to engage in the behaviour (lower part). Preferences for network rewiring show, second, that actors indeed prefer to be connected to behaviourally similar other actors (positive *behaviour similarity* effect in the network part). This preference for similar network contacts occurs on top of several empirically well-known control dimensions: Actors tend to select few network contacts (negative *outdegree* estimate) and base their decision on whether their network ties will be mutual (positive *reciprocity* effect), embedded in the same network subgroup (*transitive triplets* effect, *"friends of friends are friends"*) and in the same gender group (effect of *same sex*). Their preference for reciprocated ties is weaker when the ties are embedded in a network subgroup than when they are not (negative interaction effect *transitive reciprocated triplets*), and actors of the same sex tend to show weaker similarity-attraction than actors of opposite sexes do (non-significant but sizeable effect). These latter two results confirm hypotheses formulated by Block (2015) and Block & Grund (2014), respectively. Finally, preferences for behaviour change indicate that actors strive to become similar to their network contacts (positive *total similarity* effect), i.e., that peer influence occurs in this network, after controlling for the marginal distributions (*linear* and *quadratic shape* effects) and a main effect of gender.

**Simulation scenarios**

The two dimensions to be manipulated, *network speed* and *amplified preferences*, are now mapped to the network parameter estimates *rate of change* and *behaviour similarity*, respectively – the latter encompassing also the interaction term with *same gender*. The operationalisation of network speed is uncontroversial, as the rate of change parameter indicates the number of opportunities that an actor on average gets for changing an (outgoing[9]) network tie. The operationalisation of preference amplification here is exclusively referring to preferences due to behavioural similarity – i.e., we study *behavioural homophily*[10] amplification. On the one hand, this particular effect is important for the applied research addressed in the simulation studies below, as was elaborated above. On the other hand, it is the only effect for which an interpretation in terms of preference is straightforward (homophily = liking = preference). *Outdegree* (the effect of the number of ties an actor has) expresses resource constraints and relational costs. *Reciprocation* of friendship and *sex segregation* in adolescence may reflect social norms rather than preferences, while *transitivity* can be interpreted better in terms of opportunity (we are *more likely to meet* our friends' friends than others) than in terms of preference (I like my friends' friends better than others).

We independently manipulate the two dimensions under study while keeping the behaviour part of the estimated model constant, as well as the networking preferences reflected in the other effects (*outdegree*, *reciprocity*, *transitive triplets*, *transitive reciprocated triplets*, and *same sex*), as follows. *Network speed* is varied between zero (*rate of change* = 0) via its empirical size (on average 13.9 opportunities per actor to change an outgoing network tie, see Table 1) and twice its empirical size (27.8 such opportunities) to thrice its empirical size (41.7 such opportunities). *Preference amplification* is varied between zero (parameters are 0.0 for behaviour similarity as well as the interaction with same gender) via half its empirical size (parameters 0.58 and -0.24, respectively) and its empirical size (1.16 and -0.48, see Table 1) to twice its empirical size (2.32 and -0.96). This gives in combination 4×4=16 conditions, one of which replicates the empirical situation, to be interpreted as "offline" reference scenario, the others embed this into a two-dimensional space, covering a spectrum between diffusion models on static networks and models that depart from the offline reference into the "online" direction. These sixteen model parametrisations will be used for simulating the outcomes of network-behaviour co-evolution processes.

## Study 1: The *echo-chamberness* of network neighbourhoods

In our first simulation study, we investigate the local network neighbourhoods that actors find themselves in after experiencing the simulated network-behaviour co-evolution process, and assess how much these resemble echo chambers in each of our simulation scenarios. A well-established index measuring the degree to which neighbourhoods are behaviourally homogeneous is Moran's index of spatial autocorrelation[11], which we use for operationalising echo-chamberness. The index assumes the values 1 and -1 for perfect positive or negative autocorrelation, and a value close to zero for absence of autocorrelation. The simulations all start out from the first observation of the offline network, and we generate a distribution of 50 simulated networks for each of the 16 scenarios.

Based on the reasoning provided above when discussing the case of echo chambers, we expect to see that compared to the empirically calibrated offline scenario (i.e., empirical network speed and empirical homophily strength), the "more online" counterfactual scenarios (i.e., higher speed and/or stronger homophily) show more evidence for echo chambers (i.e., higher autocorrelation indices; *first expectation*). Furthermore, we expect that if there is more randomness in the behaviour distribution of chosen contacts (i.e., homophily preferences are not amplified but reduced compared to the reference scenario) we actually will observe the less evidence for echo chambers, the more the network is rewired (i.e., lower autocorrelation indices; *second expectation*).

**Results**

The sixteen scenarios lead to sixteen simulated distributions of network-behaviour configurations at the end of the co-evolution period. Each simulated network-behaviour configuration then is evaluated in terms of Moran's network autocorrelation index, resulting in sixteen distributions of this index, which are depicted in Figure 1.

> *Insert Figure 1 about here* <

Figure 1 shows the expected interaction pattern between the two types of rewiring that were manipulated. Depending on the strength of actors' preference for homophilous ties, the relative speed at which network changes happen can either reduce behavioural homogeneity of network neighbours (for low levels of homophily) or amplify it (for high levels of homophily). Interestingly, the empirically found strength of homophily in the third panel of the figure seems to be the tipping point at which the effect of overall rewiring intensity on neighbourhood homogeneity changes from negative to positive. In other words, the actors in the data set under study exhibit a level of homophily that keeps observed neighbourhood homogeneity intact, regardless of overall rewiring speed. We tentatively conclude that social actors in this real-life social network seem to have formed relational preferences that make their neighbourhood homogeneity *insensitive* to increased or lowered networking speeds. This is a delicate position, as a tiny tinkering with these relational preferences – e.g., a slight increase or decrease of actors' propensity to choose homophilous neighbours – will make the homogeneity of their network neighbourhood sensitive to the speed of networking. In online scenarios, even a slight amplification of homophily might therefore propel the network as a whole into a situation of segregated echo chambers.

## Study 2: The success of peer-led network interventions

Based on the above, we expect that any particular actor's influence on others is diminished when network speed increases, and that peer-led network interventions therefore are less effective (*third expectation*). When in addition, homophily preferences get amplified, we expect that neighbourhoods become so homogeneous that influence process cannot contribute much additional homogeneity, with again the result that the effects of peer-led network interventions are reduced (*fourth expectation*). With our second study, we want to deepen our understanding of these processes by checking whether this reasoning indeed bears out. We investigate whether the two dimensions we identified as major differences between online and offline, rewiring speed and preference amplification, can undermine the influence of those peers who are known to have high influence offline. We will again study this by considering a series of counterfactual scenarios that, first, help us establish the workings of a classical *peer-led intervention study* (explained directly below) in the empirical reference scenario. Second, we will check how sensitive the intervention effect is to the departures from this reference scenario described earlier.

For the first task, we identify those actors who belong to the top 15% most popular in the first observation of the network as our group of *influentials*. Their collective behaviour is then artificially fixed to one of the values on the empirical four-point scale of the behaviour (i.e., they all are made to score 1 in one scenario, they all score 2 in another scenario, etc.). In the simulations, which are conducted as in the study above, these offline influential actors are not allowed to change their behaviour, only the other actors are allowed to do this. The average behaviour of these other actors at the end of the simulation period then is the outcome further investigated. Intuitively, the higher the group of influentials is made to score, the higher also the influenced remaining actors should score at the end of the period. And indeed, this is what Figure 2 indicates.

> *Insert Figure 2 about here* <

We can now run a linear regression of influenced actors' behaviour on influential actors' behaviour. The slope of this regression line (also depicted in Figure 2) indicates how sensitive the influenced's behaviour is to the influentials'. In other words, the slope is an indicator of the *success of the peer-led network intervention* that we simulated. Noting that this scenario makes use of the empirically obtained estimates of the co-evolution model, we now run analogous studies also for the 15 counterfactual scenarios, which exhibit various combinations of network speed and preference amplification, and obtain a regression slope for each of these scenarios.

**Results**

Figure 3 shows how the success of the intervention depends on network speed and preference amplification (again in the shape of behavioural homophily amplification).

> *Insert Figure 3 about here* <

What we see is first of all that network speed and preference amplification have the expected negative effects on intervention success. Furthermore, we see that the assumption of a static network (zero speed) leads to empirically unrealistic predictions of intervention success. As argued above, we suspect that many researchers designing and conducting such peer-led network interventions may indeed have had such unrealistic, high expectations because it is very common in network science to assume static networks on which diffusion processes unfold. In our view, this explains why with their chosen study design, they could not detect an effect of their interventions: assuming an inflated effect and a static network, they may have chosen too small a sample to be able to detect the actually much weaker effect under a dynamic network.

## Discussion

The two counterfactual but empirically anchored studies reported in this research note illustrate how the *rewiring speed* of social networks and the rewiring pattern of *homophily* (McPherson et al., 2001) can have strong effects on system-level properties of the network. Study 1 showed how segregation into echo chambers emerges under conditions of high network speed and amplified levels of homophily, and how echo chambers dissolve under conditions of high network speed but reduced levels of homophily. Study 2 showed that interpersonal influence can be exploited best for promoting behaviour change under conditions of a slow network that tends not to self-organise according to behavioural homophily. Both studies highlight that exactly the same peer influence mechanism operating in a social network can produce vastly different outcomes, depending on the details of the network change process that occurs in parallel to the influence process, simply because it largely determines the input of peer influence. These emergent phenomena, and their consequences in real life, are so far insufficiently understood in the community of network scholars, let alone the more applied disciplines. Cases in point are two different research domains that we address in our article.

### Conclusions from Study 1

Based on actor-level analysis of online data, some researchers have argued that while actors' network decisions were significantly affected by algorithmic amplification of preferences, this effect is very small compared to the main effect of these preferences (Bakshy et al., 2015; Benkler et al., 2017; Brühl et al., 2017). In the light of the results of Study 1, we think a word of caution needs to be added to this assent to personalisation algorithms. Our results suggest that even small amplifications of preference can spiral the online situation into a quite segregated universe of echo chambers, due to the internet's high speed of offering new network contacts. There seems to be a delicate balance in offline situations between how humans adjust their behaviour to the input they receive from network neighbours, and how they adjust the composition of their network neighbourhood to the network rewiring opportunities they have. If the latter process speeds up, even small tinkering with the rewiring preferences can render the former process unimportant (in the case of preference amplification) or highly important (in the case of preference reduction). Personalisation algorithms that bring us into contact with always more of the same will make the online experience dull and pointless, as we cannot learn anything new anymore, *even if we learn*, and we get stuck in homogenous echo chambers. By contrast, if personalisation brings us into contact with more variability than what our own preferences would produce, we might actually be pulled out of our self-inflicted echo chambers and broaden our minds to tolerate a larger spectrum of attitudes and opinions out there.

**Conclusions from Study 2**

Deeper inquiry into the mechanisms producing echo chambers led us to a closer inspection of the workings of peer influence. Researchers trying to implement peer-led network interventions have suffered from highly divergent and often smallish success rates of these interventions (Kim & Free, 2008; McArthur et al., 2016). We think we can explain this heterogeneity based on the results of our second study. When assuming a network to be static, and in particular: not dynamically responding to the diffusion process that is taking place on it, then this can lead to an overestimation of the expected effect of the intervention, and ultimately to disappointed researchers when finding only a much smaller than anticipated effect. This assumption of a static network (of varying topology) as the substrate on which influence processes unfold is quite common in network studies of diffusion processes (e.g., Barzel & Barabási 2013; Centola, 2010; Delre et al., 2010; Pastor-Satorras & Vespignani, 2003; Watts & Dodds, 2007) and may have tricked some applied intervention practitioners into believing that only network topology, and the intervention targets' position in this topology, mattered for the success of the intervention, but not dynamic rewiring of this topology. Our Study 2 shows that this rewiring, which determines who the peers are that deliver the input for any peer influence process, has a profound impact on the intervention's success.

Taken together, our twin studies illustrate how a network science perspective that abstracts from the concrete social science content of network research is able to give a more complete understanding of the principles according to which networks tend to self-organise. We could meaningfully address puzzles from two quite different research fields by a common approach, and hopefully could contribute to a better understanding of the investigated phenomena in both disciplines.

**Limitations & follow-up**

It is clear that internet companies keep the details of their personalisation algorithms secret. Our operationalisation of personalisation as a simple preference amplification for homophily in person-to-person networks does not do justice to the complexities of the machine learning algorithms that underlie actual personalisation algorithms. We think that our operationalisation is a reasonable approximation of what may go on, but would certainly agree that other operationalisations should be explored in studies like ours, too. On the one hand, this refers to the network type. We now focused on person-to-person networks, but likewise or additionally, person-to-product or person-to-media-content networks should be studied with our approach. On the other hand, different recommender algorithms should be investigated – e.g., the well-known product recommender by closing bipartite 4-cycles "you have bought X – people who bought X also bought Y – won't you, too, want to buy Y?", or the transitivity recommender "Z is a friend of your friend – won't you have Z as friend, too?" Both of these do not explicitly refer to homophily being manipulated, but because they are clustering mechanisms, they are likely to amplify existing homophily in users' preferences (Goodreau et al., 2009; Steglich & Knecht, 2010) and, though more indirectly, may lead to the same qualitative conclusions as the present study does.

Furthermore, it is well-known that popularity tends to be self-reinforcing in online contexts (known as *cumulative advantage*, De Solla Price, 1965, 1976; the *rich-get-richer* or *Matthew effect*, Merton, 1968; or *preferential attachment*, Barabási & Albert, 1999), so part of the reasons we discussed why an offline network intervention can fail may actually not carry over into the online world because there, popular actors tend to stay popular. Given this, the value of the second simulation study for the internet argument we are trying to present remains unclear. A follow-up study including popularity-preserving mechanisms in the simulation models would probably show that the negative effects of rewiring may be less severe than in the current study, but it is hard to believe they would entirely vanish due to the remaining randomness in the rewiring algorithm.

Last but not least, the simulations reported in this paper are anchored at a single empirical dataset, which potentially could limit the generalisability of our findings. In particular, the diagnosis of a balance between adjustment to peers and replacement of peers depends on this balance being present in the empirical data. We consider this danger to be small because the dataset is fairly representative for offline contact networks on this and many other descriptive network dimensions.

**The larger picture**

For understanding the societal consequences of online-facilitated connectedness, it is important to consider humans' limited information processing capacity. As we could show, even when the rules of individual behaviour are not directly affected by online algorithms, there are profound indirect effects[12]. The networks in which individuals operate online change rapidly and are affected by personalisation algorithms under control of the providers of network infrastructure. To blame the negative side effects of online infrastructure – like the emergence of echo chambers online – on individuals' decisions, and downplay the role of the algorithms (Bakshy et al., 2015) is a flawed, incomplete and dangerous conclusion. These individual decisions take place in a highly pre-structured environment, the structuring of this environment is done by said algorithms, and as we could show, the nature of this algorithmic foundation of individuals' internet experience pretty much pre-determines the collective outcome of the decisions.

Our freedom is supposed to manifest itself in the decisions we make. This includes in particular decisions about the social relationships we form and maintain, but also in decisions what to buy or what news sources to pay attention to. These decisions are embedded in feedback loops: What we choose today can limit or widen our horizon of future opportunities, thus having strong impact on our future freedom. Of special importance in this context are relationships with those people who are not members of our own group, and exposure to opinions and information that we may initially disagree with. Granovetter (1973) famously argued that the strength of weak ties lies in their potential to give us access to *different* information, i.e., information that is so far unknown to us. And what are online contacts if not weak ties? There is a lot of potential in online networks for bringing us into contact with the wide world out there. However, the strength of weak ties is lost when online recommender systems rely on our past expressed preferences for suggesting to us new contacts, new products, or new sources of information. On the contrary, this bears the danger of actually sealing us off this wide world and separating us into echo chambers online (Study 1). Because these echo chambers may be so large that we do not experience them as a limit on our freedom, because we therefore might actually mistake them for the whole world, there is a danger that personalisation algorithms might facilitate the emergence of *parallel societies* without anyone being aware of it[13].

As Study 2 showed, compared to classical forms of offline peer influence, the online situation can facilitate a massive shift away from sources that offline would be highly influential, towards sources that recommender algorithms deem appropriate. It is clear that such disempowerment of individuals should not entirely be guided by commercial concerns of the providers of online infrastructure, such as their desire to "match", i.e., deliver the most-likely-to-buy customers to their commercial clients. To demand *transparency* in algorithms may be a solution, but probably not a very realistic one. Anecdotally, my son once said, in the earlier days of Facebook, *"Only a complete fool would think a Facebook friend is a real friend."* Hiding in this statement is an imperative to differentiate, and not to apply the same rules and heuristics for handling online content and contacts as for handling offline content and contacts. A more feasible task than asking the internet companies for transparency in their algorithms may therefore be the task to *educate* people better about the out-of-sight consequences of online infrastructure. Further investigative research about the pitfalls of online infrastructure[14] is much needed and has to be included in educative programs on *digital citizenship* and in particular *internet* and *social media literacy* (e.g., Lankshear & Knobel, 2008). We maybe cannot make the giant internet companies change their algorithms for their own economic worse, but they will change their algorithms when we as users devise ways to appreciate more diverse content.

## References

Bakshy, E., Messing, S., & Adamic, L. A. (2015). Exposure to ideologically diverse news and opinion on Facebook. *Science*, *348*(6239), 1130-1132.

Barabási, A. L., & Albert, R. (1999). Emergence of scaling in random networks. *Science*, *286*(5439), 509-512.

Barzel, B., & Barabási, A. L. (2013). Universality in network dynamics. *Nature physics*, *9*(10), 673-681.

Benkler, Y., Faris, R., Roberts, H., & Zuckerman, E. (2017). Study: Breitbart-led right-wing media ecosystem altered broader media agenda. *Columbia Journalism Review*, *1*(4.1), 7.

Block, P. (2015). Reciprocity, transitivity, and the mysterious three-cycle. *Social Networks*, *40*, 163-173.

Block, P., & Grund, T. (2014). Multidimensional homophily in friendship networks. *Network Science*, *2*(02), 189-212.

Boxell, L., Gentzkow, M., & Shapiro, J. M. (2017). *Is the internet causing political polarization? Evidence from demographics* (No. w23258). National Bureau of Economic Research.

Brandes, U., Robins, G., McCranie, A., & Wasserman, S. (2013). What is network science? *Network Science*, *1*(01), 1-15.

Brühl, J., Brunner, K., & Ebitsch, S. (2017). Der Facebook-Faktor. Wie das soziale Netzwerk die Wahl beeinflusst. *Süddeutsche Zeitung*, 3 May 2017.

Campbell, R., Starkey, F., Holliday, J., Audrey, S., Bloor, M., Parry-Langdon, N.,…Moore, L. (2008). An informal school-based peer-led intervention for smoking prevention in adolescence (ASSIST): a cluster randomised trial. *The Lancet*, 371, 1595-1602.

Centola, D. (2010). The spread of behavior in an online social network experiment. *Science*, *329*(5996), 1194-1197.

Davenport, T. H., & Beck, J. C. (2001). *The attention economy: Understanding the new currency of business*. Harvard Business Press.

De Solla Price, D. J. (1965). Networks of scientific papers. *Science*, 510-515.

De Solla Price, D. J. (1976). A general theory of bibliometric and other cumulative advantage processes. *Journal of the Association for Information Science and Technology*, *27*(5), 292-306.

Delre, S. A., Jager, W., Bijmolt, T. H., & Janssen, M. A. (2010). Will it spread or not? The effects of social influences and network topology on innovation diffusion. *Journal of Product Innovation Management*, *27*(2), 267-282.

DiMaggio, P., Hargittai, E., Celeste, C., Shafer, S. (2004). Digital inequality: from unequal access to differentiated use. In Neckman, K. (ed.), *Social Inequality*. New York: Russell Sage. pp. 355-400.

Dunbar, R. I. (1998). The social brain hypothesis. *brain*, *9*(10), 178-190.

Epstein, J. M., Parker, J., Cummings, D., & Hammond, R. A. (2008). Coupled contagion dynamics of fear and disease: mathematical and computational explorations. *PLoS ONE*, *3*(12), e3955.

Farrell, H. (2012). The consequences of the internet for politics. *Annual review of political science*, *15*, 35-52.

Goodreau, S. M., Kitts, J. A., & Morris, M. (2009). Birds of a feather, or friend of a friend? Using exponential random graph models to investigate adolescent social networks. *Demography*, *46*(1), 103-125.

Granovetter, M. S. (1973). The strength of weak ties. *American journal of sociology*, *78*(6), 1360-1380.

Homans, G. C. (1950). *The human group*. New York: Harpers.

Kelly, J. A., & Stevenson, L. Y. (1995). Opinion leader HIV prevention training manual. *Milwaukee, WI: Center for AIDS Intervention Research, Medical College of Wisconsin*.

Kim, C. R., & Free, C. (2008). Recent evaluations of the peer-led approach in adolescent sexual health education: A systematic review. *Perspectives on sexual and reproductive health*, *40*(3), 144-151.

Lakon, C. M., Hipp, J. R., Wang, C., Butts, C. T., & Jose, R. (2015). Simulating dynamic network models and adolescent smoking: the impact of varying peer influence and peer selection. *American journal of public health*, *105*(12), 2438-2448.

Lankshear, C., & Knobel, M. (2008). *Digital literacies: Concepts, policies and practices* (Vol. 30). Peter Lang.

Latkin, C. A. (1998). Outreach in natural settings: the use of peer leaders for HIV prevention among injecting drug users' networks. *Public Health Reports*, *113*(Suppl 1), 151.

Lazarsfeld, P. F., & Merton, R. K. (1954). Friendship as a social process: A substantive and methodological analysis. *Freedom and control in modern society*, *18*(1), 18-66.

Leskovec, J., Adamic, L. A., & Huberman, B. A. (2007). The dynamics of viral marketing. *ACM Transactions on the Web (TWEB)*, *1*(1), 5.

McArthur, G. J., Harrison, S., Caldwell, D. M., Hickman, M., & Campbell, R. (2016). Peer-led interventions to prevent tobacco, alcohol and/or drug use among young people aged 11–21 years: a systematic review and meta-analysis. *Addiction*, *111*(3), 391-407.

McPherson, M., Smith-Lovin, L., & Cook, J. M. (2001). Birds of a feather: Homophily in social networks. *Annual review of sociology*, *27*(1), 415-444.

Mercken, L., Sleddens, E. F., de Vries, H., & Steglich, C. (2013). Choosing adolescent smokers as friends: The role of parenting and parental smoking. *Journal of adolescence*, *36*(2), 383-392.

Mercken, L., Steglich, C., Sinclair, P., Holliday, J., & Moore, L. (2012). A longitudinal social network analysis of peer influence, peer selection and smoking behaviour among adolescents in British schools, *Health Psychology* 31, 450-459.

Merton, R. K. (1968). The Matthew effect in science. *Science*, *159*(3810), 56-63.

Moran, P. A. (1948). The interpretation of statistical maps. *Journal of the Royal Statistical Society. Series B (Methodological)*, *10*(2), 243-251.

Osgood, D. W., Ragan, D. T., Wallace, L., Gest, S. D., Feinberg, M. E., & Moody, J. (2013). Peers and the emergence of alcohol use: Influence and selection processes in adolescent friendship networks. *Journal of Research on Adolescence*, *23*(3), 500-512.

Pariser, E. (2011). The Filter Bubble: What the Internet Is Hiding from You. New York: Penguin Press.

Pariser, E. (2015). Did Facebook's Big Study Kill My Filter Bubble Thesis? *Wired*, 7 May 2015.

Pastor-Satorras, R., & Vespignani, A. (2003). Epidemics and immunization in scale-free networks. In *Bornholdt, S., & Schuster, H. G. (Eds.), Handbook of Graphs and Networks: From the Genome to the Internet*, 111-130.

Qiu, X., Oliveira, D. F., Shirazi, A. S., Flammini, A., & Menczer, F. (2017). Limited individual attention and online virality of low-quality information. *Nature Human Behavior*, *1*, 0132.

Sandvig, C. (2015). The Facebook "It's Not Our Fault" Study. *Social Media Collective* research blog. URL: https://socialmediacollective.org/2015/05/07/

Schaefer, D. R., Adams, J., & Haas, S. A. (2013). Social networks and smoking: exploring the effects of peer influence and smoker popularity through simulations. *Health Education & Behavior*, *40*(1_suppl), 24S-32S.

Smit, C. R., de Leeuw, R. N., Bevelander, K. E., Burk, W. J., & Buijzen, M. (2016). A social network-based intervention stimulating peer influence on children's self-reported water consumption: A randomized control trial. *Appetite*, *103*, 294-301.

Snijders, T. A. (1996). Stochastic actor-oriented models for network change. *Journal of mathematical sociology*, *21*(1-2), 149-172.

Snijders, T. A. (2001). The statistical evaluation of social network dynamics. *Sociological methodology*, *31*(1), 361-395.

Snijders, T. A., & Steglich, C. (2015). Representing micro–macro linkages by actor-based dynamic network models. *Sociological methods & research*, *44*(2), 222-271.

Snijders, T., Steglich, C., & Schweinberger, M. (2007). Modeling the Co-Evolution of Networks and Behavior, in K. van Montfort, H. Oud & A. Satorra (Eds.), *Longitudinal models in the behavioral and related sciences*. Mahwah NJ: Lawrence Erlbaum, pp. 41-71.

Snijders, T. A., Van de Bunt, G. G., & Steglich, C. (2010). Introduction to stochastic actor-based models for network dynamics. *Social networks*, *32*(1), 44-60.

Starkey, F., Audrey, S., Holliday, J., Moore, L., & Campbell, R. (2009). Identifying influential young people to undertake effective peer-led health promotion: the example of A Stop Smoking in Schools Trial (ASSIST). *Health Education Research*, 24, 977-988.

Steglich, C. (2007). Closure, constraint and homophily: Joint determinants of network segregation. *Unpublished working paper, University of Groningen*. DOI: 10.13140/RG.2.1.4961.7442.

Steglich, C., & Knecht, A. (2010). Die statistische Analyse dynamischer Netzwerke. *Handbuch Netzwerkforschung*, 433-446.

Steglich, C., Snijders, T. A., & Pearson, M. (2010). Dynamic networks and behavior: Separating selection from influence. *Sociological methodology*, *40*(1), 329-393.

Steglich, C., Snijders, T. A. B., & West, P. (2006). Applying SIENA: An illustrative analysis of the coevolution of adolescents' friendship networks, taste in music, and alcohol consumption. *Methodology*, *2*(1), 48-56.

Sunstein, C. R. (2001). *Echo chambers: Bush v. Gore, impeachment, and beyond*. Princeton, NJ: Princeton University Press.

Tufekci, Z. (2013). "Not This One" Social Movements, the Attention Economy, and Microcelebrity Networked Activism. *American Behavioral Scientist*, 57(7), 848-870.

Tufekci, Z. (2015). Facebook said its algorithms do help form echo chambers, and the tech press missed it. *New Perspectives Quarterly*, *32*(3), 9-12.

Valente, T. W. (2005). Network models and methods for studying the diffusion of innovations. In Carrington, P., Scott, J. & Wasserman, S. (Eds.), *Models and methods in social network analysis*. Cambridge MA: Cambridge University Press. pp. 98-116.

Valente, T. W. (2012). Network interventions. *Science*, *337*(6090), 49-53.

Valente, T. W., & Pumpuang, P. (2007). Identifying opinion leaders to promote behavior change. Health Education & Behavior, 34, 881-896.

Watts, D. J., & Dodds, P. S. (2007). Influentials, networks, and public opinion formation. *Journal of consumer research*, *34*(4), 441-458.

Zollo, F., Bessi, A., Del Vicario, M., Scala, A., Caldarelli, G., Shekhtman, L., ... & Quattrociocchi, W. (2017). Debunking in a world of tribes. *PLoS one*, *12*(7), e0181821.

---

[1] It is immaterial whether we here follow Robin Dunbar's *social brain* hypothesis (Dunbar, 1998) and set the limit to 150, or more broadly reference empirical results from social network analysis which identified … contact persons. Whatever the details are, the crucial fact here is the existence of cognitive limits (in the study reported in this article, we limit the number of network contacts to a maximum of six).

[2] The aim behind these personalisation efforts is the matching of users with content, an amalgam of service mindedness and commercial interest of the companies providing networking infrastructure. On the one hand, such matching aims to maximise relevance of the online experience to the individual user, on the other hand, it is to maximise relevance of online advertising targets to the commercial clients.

[3] In the study reported here, we disregard the possibility that repeated assessment of preferences from an already filtered set of opportunities might actually still further aggravate this amplification.

[4] We write, "current", because personalisation of the kind described *currently* is a feature of search engines and social networking sites, but it has not been one in the past, and could be abandoned again in the future.

[5] This reasoning has led researchers to study selection processes as *precursors* of peer influence, e.g., in determining which adolescents pick friends who will subsequently expose them to risk behaviour (Mercken et al., 2013; Osgood et al., 2013).

6  Similar work for peer influence *online* can be found under the keyword *viral marketing* (e.g., Leskovec et al., 2007), a marketing strategy that attempts to exploit influence processes in online networks.

7  From a social science viewpoint, it is hard to understand why in the network science literature so many network diffusion studies are assuming static networks. Starting out with the assumption that people strive for similarity to their peers when they decide about own opinions or behavior, it seems hard to justify that they would not express the same preferences when deciding about whom to be connected to.

8  The term *behaviour* must not be taken literally here, but stands shorthand for *changeable* characteristic of the network actors, i.e., encompasses also individual attitudes, opinions, health outcomes, *et cetera*.

9  In the stochastic actor-based modelling framework of directed networks, an actor's *outgoing* ties are assumed to be *under the control* of the actor (e.g., whom to trust, whom to call a friend, etc.) while incoming ties are not (I cannot decide who trusts me, who calls me a friend, *etc*.).

10  Recognising divergent use in various research fields, as sociologists we stay true to Lazarsfeld & Merton's (1954) original definition of the word *homophily* as 'love of like', i.e., as preference for similar others that is expressed in decisions about whom to be in contact with.

11  Moran's autocorrelation index (Moran, 1948) is defined as $\frac{n \sum_{i,j=1}^{n} x_{ij}(z_i - \bar{z})(z_j - \bar{z})}{\sum_{i,j=1}^{n} x_{ij} \sum_{i=1}^{n}(z_i - \bar{z})^2}$, where $n$ is the size of the network (i.e., the number of network members), $x$ stands for the matrix of connectivity strengths (in our case binary, indicating contact), and $z$ stands for the actor-level variable for which network autocorrelation is quantified. The expected value of the index (under random allocation of $z$ preserving its marginal distribution) is not exactly zero, but $1/(1-n)$.

12  In a similar vein, also Qiu et al. (2017) recently could show that such cognitive limits facilitate the spread of low-quality information (e.g., fake news) in the internet-typical situation of information overload.

13  Zollo et al. (2017) in this context speak of a "world of tribes", and actually find evidence for it in their study on the persistence of fake news in Facebook groups.

14  We think of initiatives such as the *Digital Polarization Initiative* (http://digipo.io/) or *AlgorithmWatch* (https://algorithmwatch.org).

> In one extreme situation, an actor's personal network which is rewired very quickly and completely at random (i.e., according to what might be called "de-personalised" rewiring) will over time expose individuals to the whole spectrum of other actors and their individual attributes (opinions, behaviours, etc.). Processes of peer influence and mutual adjustment here are likely to produce a global consensus. However, in another extreme situation, a network rewired very quickly but completely based on past preferences (i.e., according to strongly personalised rewiring) will limit the exposure of individuals to actors who provide very similar input. In this situation, the same peer influence processes are likely to not only reproduce existing group differences, but even exacerbate them, and lead to increased global polarisation.

**Textbox 1.**

> McArthur et al. (2016) reviewed the existing RCT studies of peer-led network interventions to reduce substance use among adolescents. Based on meta-analyses, they concluded that while these interventions generally did work (i.e., there was a significant overall mean substance use reduction in the intervention groups compared to the control groups), effects differed strongly between the reviewed studies, and the overall mean effect was very weak. A similar review article by Kim & Free (2008) on interventions promoting safe sex practices also found strong heterogeneity of effects in individual studies, but no systematic overall effect.

**Textbox 2.**

| NETWORK PART | estimate | st.err. | sig. |
|---|---|---|---|
| rate of change | 13.9 | 1.04 | |
| outdegree | -3.49 | 0.13 | *** |
| reciprocity | 2.57 | 0.15 | *** |
| transitive triplets | 0.83 | 0.06 | *** |
| trans. recip. triplets | -0.63 | 0.10 | *** |
| same sex | 0.78 | 0.13 | *** |
| behaviour similarity | 1.16 | 0.62 | † |
| same sex × behaviour similarity | -0.48 | 0.63 | |
| | | | |
| BEHAVIOUR PART | estimate | st.err. | sig. |
| rate of change | 5.41 | 2.87 | |
| linear shape | -1.56 | 0.43 | *** |
| quadratic shape | 0.87 | 0.10 | *** |
| total similarity | 0.84 | 0.38 | * |
| female | 0.08 | 0.24 | |

† p<0.1; * p<0.05; ** p<0.01; *** p<0.001

**Table 1.** Empirical estimates obtained from a stochastic actor-based analysis of network-behaviour co-evolution data. These results are modified in the simulation studies.

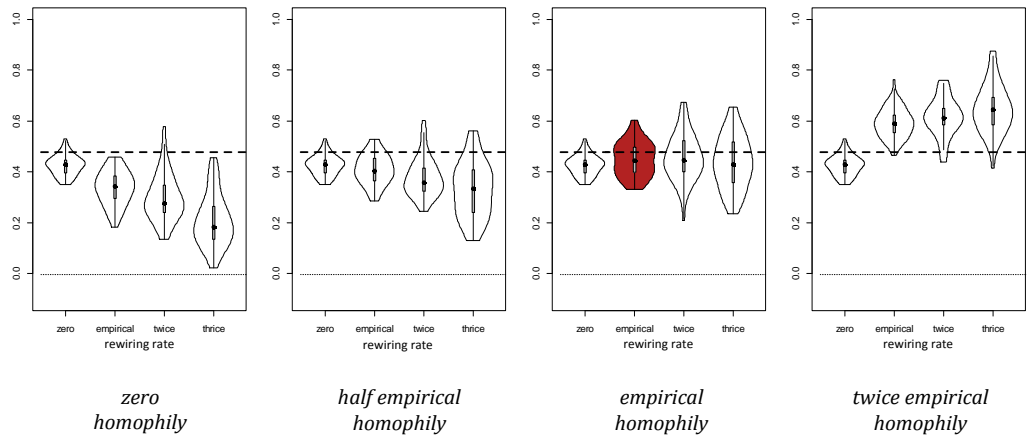

**Figure 1.** Sensitivity of Moran's autocorrelation index (y-axis) to network speed (x-axis) and strength of behaviour homophily (panels left to right). The dashed and dotted lines indicate, respectively, the empirically observed autocorrelation value and the value expected under complete randomness. Results for the empirically obtained parameter set are marked red.

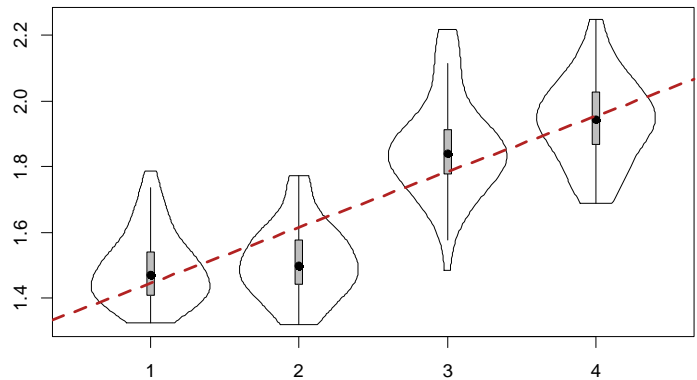

**Figure 2.** Distribution of non-influential actors' average behaviour (vertical axis) depending on influential actors' behaviour (manipulated, horizontal axis), for the empirical scenario shown in Table 1. The slope of the regression line indicates success of the intervention.

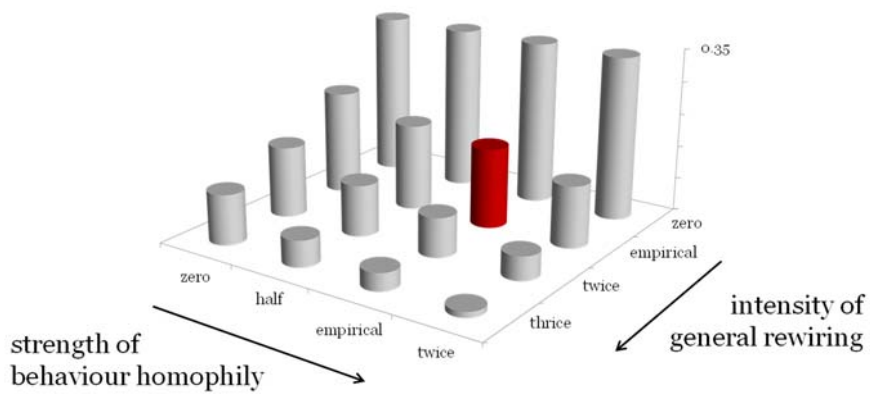

**Figure 3.** Sensitivity of intervention success (i.e., slopes of the type depicted in Figure 2) to general rewiring intensity (network speed) and strength of specific rewiring in the form of behaviour homophily. Results for the empirically obtained parameter set are marked red.